%% file: samplepaper.tex
\documentclass[runningheads]{llncs}
\usepackage{graphicx}
\begin{document}
%
\title{Machine Learning and value generation in Software Development: a survey}
\titlerunning{ML applications for Software Development}
%
\author{
Barakat. J. Akinsanya
\and Luiz J.P. Ara\'ujo \inst{1} 
\and Mariia Charikova
\and Susanna Gimaeva
\and Alexandr Grichshenko 
\and Adil Khan
\and Manuel Mazzara 
\and Ozioma Okonicha N
\and Daniil Shilintsev}

\authorrunning{Akinsanya et al.}

%
\institute{Innopolis University, Innopolis 420500, Russia\\
\inst{1} Corresponding author: \email{l.araujo@innopolis.university}
}
\maketitle              
\vspace{-5 mm}
\begin{abstract}
Machine Learning (ML) has become a ubiquitous tool for predicting and classifying data and has found application in several problem domains, including Software Development (SD). This paper reviews the literature between 2000 and 2019 on the use the learning models that have been employed for programming effort estimation, predicting risks and identifying and detecting defects. This work is meant to serve as a starting point for practitioners willing to add ML to their software development toolbox. It categorises recent literature and identifies trends and limitations. The survey shows as some authors have agreed that industrial applications of ML for SD have not been as popular as the reported results would suggest. The conducted investigation shows that, despite having promising findings for a variety of SD tasks, most of the studies yield vague results, in part due to the lack of comprehensive datasets in this problem domain. The paper ends  with concluding remarks and suggestions for future research. 
\end{abstract}
\vspace{-5 mm}

\input{1_Introduction.tex}
\vspace{-5 mm}
\input{2_Effort.tex}
\vspace{-5 mm}
\input{3_Risks.tex}
\vspace{-5 mm}
\input{4_Defects.tex}
\vspace{-5 mm}
\input{5_Discussion.tex}
\vspace{-5 mm}
\input{6_Conclusion.tex}
\vspace{-5 mm}





\begin{table}[h]
\caption{Machine Learning for Software Development in academic literature.}
\label{reference table}
\begin{tabular}{|p{2.7cm}|c|p{4.5cm}|p{3.5cm}|}
\hline
Reference & Task & ML model & Data\\
\hline
Azzeh (2011) & SEE\textsuperscript{1} & Decision Tree & PROMISE and ISBSG datasets \\ \hline

Bardsiri and Hashemi (2017) & SEE\textsuperscript{1} & Regression Trees, ANN & ISBSG and NASA datasets \\ \hline

Baskeles, Turhan, and Bener (2007) & SEE\textsuperscript{1} & Multilayer Perception, Regression Trees, Support Vector Regression & NASA and USC datasets \\ \hline

Catal, Diri, and Ozumut (2007) & SFP\textsuperscript{2} & Artificial Immune Systems paradigm & PROMISE dataset \\ \hline

Ceylan, Kutlubay, and Bener (2006) & SFP\textsuperscript{2} & Decision Trees, Multilayer Perception, Radial Basis Functions & NASA dataset \\ \hline

Clemente, Jaafar, and Malik (2018) & 
SFP\textsuperscript{2} & ANN, Random Forest, Decision Trees, Naive Bayes, SVM & SeaMonkey, Mozilla Firefox \\ \hline

Dragicevic, Celar, and Turic (2017) & SEE\textsuperscript{1} & Bayesian Network & Historical data\\ \hline

Hu et al. (2007) & SRP\textsuperscript{3} & ANN, Support Vector Machine & Questionnaire based data\\ \hline

Joseph (2015) & SRP\textsuperscript{3} & ANN & Oracle dataset \\ \hline

Karim et al. (2017) & SFP\textsuperscript{2} & SVM, ANN, Naive Bayes, Random Forest & PROMISE dataset\\ \hline

Kim and Lee (2005) & SEE\textsuperscript{1} & ANN, Regression Tree & ISBSG dataset\\ \hline

Marian et al. (2016) & SFP\textsuperscript{2} & Fuzzy decision tree & JEdit(version4.2), Ant(version 1.7)\\ \hline

Moharreri et al. (2016) & SEE\textsuperscript{1} & Decision Trees, Random Forest, Logistic Model Tree, Naive Bayes & IBM Rational Team Concert data\\ \hline

Nassif et al. (2016) & SEE\textsuperscript{1} & ANN & ISBSG dataset \\ \hline

Panda,Satapathy, and Rath (2015) & SEE\textsuperscript{1} & ANN & Zia dataset\\ \hline

Perkusich et al. (2015) & SFP\textsuperscript{2} & Bayesian Networks & Case studies in software companies\\ \hline

Ren et al. (2014) & SFP\textsuperscript{2} & Partial Least Squares and Kernel principal component analysis & NASA and SOFTLAB datasets \\ \hline

Sharma and Singh (2017) & SEE\textsuperscript{1} & ANN, Fuzzy logic, Genetic Algorithms, Regression Trees & NASA, ISBSG, Desharnais and COCOMO datasets. \\ \hline

Shepperd and Schofield (1997) & SEE\textsuperscript{1} & Case-Based Reasoning & Albrecht, Atkinson, Desharnais, Finnish and MM2 datasets\\ \hline

Wright and Ziegler (2019) & SEE\textsuperscript{1} & Neural Hidden Markov Model, Deep Mixture Density Networks & LGTM dataset\\ \hline
\end{tabular}
\newline
\textsuperscript{1}SEE: Software Effort Estimation \newline
\textsuperscript{2}SFP: Software Fault Prediction \newline
\textsuperscript{3}SRP: Software Risks Prediction
\end{table}

\input{bibliography.tex}

%




\end{document}

%% file: 1_Introduction.tex
\section{Introduction}

Software has become an essential part of modern everyday life and has a ubiquitous presence in diverse sectors including manufacturing, agriculture and health industries, to mention a few \cite{casale2016current}. Efficient software development is, therefore, essential for organisations and requires proper planning and execution in order to generate high quality software at appropriate time and cost. There are several activities involved in this developmental process of a software such as coding, testing and management of the software development cycle. Not surprisingly, issues may arise during the software life-cycle including underestimation of necessary programming effort, poor code and external aspects that implicate in risks to the project \cite{srinivasan1995machine}.
These challenges hinder the growth of businesses since it is 
considered the top priority for most organisations. The prediction, mitigation and identification of response actions to issues during software development are complex tasks often performed by human agents who use information and employ subjective expertise \cite{jorgensen2006systematic}.
The support and automatising of such tasks has gained increasing attention in the literature. Researchers over the years have produced different ideas to enhance software development by introducing statistical and regressional models.
Some of the prevalent statistical models used for this purpose include Bayesian networks \cite{perkusich2015procedure}, fuzzy logic \cite{engel2007modeling} and system dynamics and discrete event simulation-based models \cite{uzzafer2013simulation}.

The use of machine learning (ML) techniques has become increasingly popular in the context of software development \cite{sharma2017systematic}.
ML is a subfield of artificial intelligence (AI) in which mathematical models identify patterns in the input data and reach a conclusion judging by the data. Thus, such algorithms can learn some information from the input (training data) and afterwards predict the answer for new data (test data). ML techniques include supervised learning, an approach characterised by the existence of prior knowledge of the input-output mapping for a training set; unsupervised learning, which algorithms proceed with no labelled data, and reinforcement learning (reward-based approach) \cite{lison2015introduction}. There are two tasks supervised learning handles: regression (predicting a continuous numerical value) and classification (assigning a label to an item). As it will be indicated in the survey, supervised learning algorithms can be employed to generate software value both for customers and developers.   

Software development is a very complicated process which includes many non-obvious things to consider when developing products. Reducing the number of software failures is one of the most challenging problems of software production. This survey aims to investigate different approaches and applications for the use of ML to software development process.

The remaining of this paper is summarised as follows. Section \ref{sec:programming-effort} presents the main ML techniques employed for predicting and estimating programming effort. Section \ref{sec:risks-project} shows how these techniques can be used to mitigate risks to the software project. Identifying software defects is performed in the Defects Section (Section \ref{sec:defects}). A discussion of the main findings from studies on ML embedded into software development processes is presented in Section \ref{sec:discussion}. Suggestions for future work is shown in Section \ref {sec:conclusion}.





%% file: 2_Effort.tex
\section{Predicting programming effort}
\label{sec:programming-effort}

Software effort estimation has received attention since the late 1970s and has been noticed to affect the workflow of the project and its overall success significantly. Moreover, programming effort underestimation often leads to missed deadlines and deterioration of the software quality; effort overestimation, on the other hand, is one of the reasons for project deceleration caused by budget constraints \cite{molokken2005comparison}.Many software effort estimation methods have been proposed to accurately estimate effort as a function of a large number of factors. The most widely employed methods \cite{sharma2017systematic} include expert models and logical statistical models (parametric models SLIM, COCOMO; regression analysis), traditional machine learning algorithms (Fuzzy Logic, Genetic Algorithms and Regression Trees) and Artificial Neural Networks. According to \cite{sharma2017systematic}, the coding effort is most often estimated in lines of code (LOC), function points (FP) \cite{dave2014neural}; use case points (UCP) \cite{ajitha2010neural} or in labour hours \cite{wright2019standard}. This section depicts the most common approaches for software development effort estimation (SDEE) in the literature, as well as their characteristics.

The importance of accurate effort predictions and the demand for automation of the estimation process have motivated the researchers to propose first parametric models in the early 80s. These models were then tested on the software datasets comprised from the real industrial data of completed projects \cite{kemerer1987empirical}. According to Srinivasan and Fisher, the three most prominent models are COCOMO, SLIM and Function Points \cite{srinivasan1995machine}. COCOMO and SLIM models rely almost exclusively on source lines of code (SLOC) as a major input, while the function point approach utilises the number of transactions and other few additional processing characteristics (online updating and transaction rates). Despite being evaluated on the available historical data (COCOMO dataset), the above models have been proven to suffer from inconsistent performances due to the noisy nature of software datasets \cite{azzeh2011software}. Bayesian Networks (BN) is a statistical model used for estimating Agile development effort \cite{dragicevic2017bayesian}. Dragicevic, Celar and Turic outlined the benefits of BNs which include the capability of handling vast uncertainties caused by the shortage of relevant information, subjective nature of a number of metrics and difficulties in gathering them. \cite{dragicevic2017bayesian}.

Another common technique for predicting effort is expert estimation, which is suitable when the domain knowledge is not leveraged by the models \cite{jorgensen2004review}. Despite its popularity, expert systems exhibit considerable human bias. One example of such system is Planning Poker, a gamified baseline strategy for SDEE in Agile environments in which developers make estimates by playing numbered cards. In a study by Moharreri et al. Plannig Poker was proven to overestimate in 40\% of instances and was shown to have a very high MMRE score of 106.8\% \cite{moharreri2016cost}. Parametric models and expert systems are still widely used in industry and studies, however the need for better generalisation and overall performance has driven the researchers to apply machine learning methods \cite{srinivasan1995machine}. 

Case-based reasoning (CBR) and decision trees (DT) have been among the most effective and researched ML models for SDEE \cite{wen2012systematic}. Results of these models are highly interpretable and are recognised as superior or at least compatible with those of parametric and effort estimation models \cite{baskeles2007software}. It was also asserted by Wen et al. that CBR is more suitable than DTs for this task since it is favourable towards smaller datasets, which is one of the biggest limitations in SDEE research \cite{wen2012systematic}. It is worth mentioning that ensemble models that different methods are often used to gain an even better precision. Moharreri et al. presented experimental evidence that DT coupled with Planning Poker produce better estimations than these models do on their own \cite{moharreri2016cost}. Genetic algorithms and fuzzy logic have been used in ensemble models, primarily handling feature selection and imprecise information provided in the datasets \cite{wen2012systematic}. 

The idea of Artificial Neural Networks (ANNs, or simply NNs), a model that has proven its potential and outperformed traditional ML methods in a number of areas, was first proposed in the 1940s and inspired by biological neurons. ANNs are an attractive approach due to their remarkable computational power: an ability to learn nonlinear relations, high parallelism, noise tolerance, learning and generalisation capabilities \cite{basheer2000artificial}. The drawbacks of applying Neural Networks are as follows: a necessity of large datasets, computational expensiveness and the fact that the results are significantly less interpretable compared to traditional machine learning methods \cite{kim2005comparison}. However, there are some methods to overcome this limitation of interpretability
\cite{sundararajan2017axiomatic}.

Comparative study of techniques such as regression tree, k-nearest neighbour, regression analysis and neural networks when applied for software development effort estimation has shown neural networks’ best estimation ability \cite{kim2005comparison}. Further consideration was given to neural networks by various researchers to emphasize their superior capabilities in effort prediction \cite{dave2014neural}. Thus, neural networks based models most often provide the best effort estimation compared to traditional ML and their accuracy increases with the amount of data supplied \cite{bardsiri2017machine}. 

%% file: 3_Risks.tex
\section{Predicting risks to the project}
\label{sec:risks-project}

Several aspects can affect and abuse the software development cycle. Predicting risks is important because it helps to mitigate delays and unforeseen expenses and dangers to the project. As it was mentioned in \cite{christiansen2015prediction}, software development projects always have more risks than other management projects because it has more technical uncertainty and complexity. Most developers look for methodology to minimize the important risks to improve their management, because the risk factor affects the success or failure of any project.

Hu et al. identified the four main types of risks \cite{hu2007software}: 
schedule: the wrong schedule may break the development even at its very first stage; budget: the correct financing is a process that requires the utmost attention to avoid the risks in software development; technical: the developers trying to make changes or fixes in the unknown code will make the relatively big amount of mistakes until they get deep into the details of their task. Even if the damage of one mistake is minor, a big number of such mistakes can be a critical fact for the project; and management risks: risks which may include the bad working environment, insufficient hardware reliability, low effectiveness of the programming etc.

Wauters and Vanhouke proposed a method for continuously assessing schedule risks which uses support vector regression which reads periodic earned value management data from the project control environment, resulting in a more reliable time and cost forecasts \cite{wauters2014support}. The parameters of the Support Vector Machine have been tuned using a cross-validation and grid search procedure, after which a large computational experiment is conducted. The results showed that the Support Vector Machine Regression outperforms the currently available forecasting methods. Additionally, a robustness experiment has been set up to investigate the performance of the proposed method when the discrepancy between training and test set becomes larger.

The wrong finance distribution will later lead to the unreasonable use of the finances and overall project fault. For solving this problem and predicting risks related with budget and finances distribution Ceylan, Kutlubay and Bener employed regression techniques to detect and identify software defects budget-related \cite{ceylan2006software}. These techniques are used to identify potentially defective software and allow corrective action to be taken before software is released to the production environment. The results of the ‘initial system structure’ show that the methods have many faulty defect predictions when the entire dataset is used. When the results are considered in terms of algorithm performances, it is seen that all of the learning algorithms used in the research have similar prediction performances having similar mean square error values.

Even a small number of technical mistakes could be a critical factor for the project. In \cite{shivaji2009reducing}, machine learning classifiers have emerged as a way to predict the existence of a bugs in a change made to a source code file. The classifier is first trained on software history data, and then used to predict bugs. Large numbers of features adversely impact scalability and accuracy of the approach. This technique is applied to predict bugs in software changes, and performance of Naive Bayes and Support Vector Machine classifiers is characterized.

Management risks in software development are one of the most global type of risks, because if they exist, most of the time they present the most prominent damage. \cite{christiansen2015prediction} aimed to predict the risks in software development projects by applying multiple logistic regression. The logistic regression was used as a tool to control the software development process. The logistic regression analyses can grade and help to point out the risk factors, which were important problems in development processes. These analytic results can lead to creation and development of strategies and highlighted problems, which are important issues to manage, control and reduce the risks of error.

%% file: 4_Defects.tex
\section{Predicting defects}
\label{sec:defects}

Software fault prediction is a process which involves the use of software metrics and algorithms to detect software components prone to error. 
One of the most crucial stages of software development life cycle is the testing stage which involves the most time and effort. It is necessary to detect faults in a software early in the software development life cycle in order to reduce software testing costs. In recent years, researchers have come up with different approaches from machine learning in order to improve the effectiveness of software testing. \cite{monden2013assessing} introduces a model of software testing which uses fault prediction to estimate cost effectiveness.

In machine learning, the task of predicting which part of software prone to fault is known to be a classification task. Classification is the process in which the computer program learns from the data input given to it, alongside algorithms known as machine learning algorithms and then uses this learning to classify new observations. The idea behind these machine learning algorithms is for machines to learn and be able to predict faults in the future. In order for this learning to happen, they have to first identify the defects then classify them. In research, software metrics are put in place to help identify the faults and test the machine learning models. A lot of metrics are used, either method level metric or class level metric. Among them are: lines Of code (LOC), weighted methods for class (WMC), coupling between objects (CBO), response for class (RFC), branch count, unique operand and total operand.

In the work of \cite{catal2007artificial},  Artificial Immune Recognition System (AIRS), an immune-insprired supervised learning algorithm, was used to create a defect model based on method-level metrics and Chidamber-Kemerer metrics suite. \cite{catal2009investigating} on other research work examined nine classifiers for each of the five public NASA datasets. According to the research, naive Bayes algorithm provides the best prediction performance for small datasets, while random forest is the best prediction algorithm for large datasets. \cite{shanthini2012applying} compared four classifiers (Naive Bayes, K-star, Random Forest and SVM). Random Forest classifier showed better results for method level metric and SVM for class level metric. \cite{malhotra2012fault} used Random Forest, Adaboost, Bagging, Multilayer Perceptron, VM, Genetic Programming. Prediction models to estimate fault proneness using dataset of Open Source ``Apache POI'' (pure Java library for manipulating Microsoft documents). The best result is shown by random forest and bagging algorithm.

A notable issue in designing machine learning models for software fault detection is the imbalance of data sets. Most researchers focus on developing models which solve this imbalance either by directly influencing the data or not. \cite{ren2014software} used the Asymmetric Kernel Principal Component Analysis Classification (AKPCAC) method based off of the kernel principal component regression algorithm proposed by\cite{rosipal2001kernel} and Asymmetric Kernel Partial Least Squares Classier (AKPLSC) method. \cite{marian2016novel}. use fuzzy decision tree, a hybrid of fuzzy logic and decision tree which proves better than the decision tree approach.

In fault prediction studies, class level metric show better prediction performance compared to method level metric \cite{karim2017software}. The major machine learning algorithms used are Fuzzy Decision Trees, Random Forest, Bagging, AKPCAC, SVM, Naive Bayes, Regression Trees and K-Star. SVM and Random Forests provide best fault prediction models as SVM produces the best accuracy in detecting faults and Random Forest is known to be good for huge dataset. On the whole, a lot of research uses various software metrics and improved machine learning algorithms to detect and predict faults.

Within development philosophies, DevOps is becoming an increasingly adopted approach and attention is rising in both industry and academia giving rise to new project, conferences and training programs \cite{DEVOPS2018} \cite{MazzaraNSSU18} \cite{Bobrov_DevOps19} \cite{Bobrov_Education19}. Considering that the DevOps toolchain generates a large quantity of data allowing the extraction of information regarding the status and the evolution of a project, this domain is emerging as particularly suitable for ML applications for SD. Our team is currently working on the implementation of an a ML-based Anomaly Detection System (ADS) and we expect the research community to increasingly focus on this aspect.

%% file: 5_Discussion.tex
\section{Discussion}
\label{sec:discussion}

Machine learning techniques have been consistently used in the last decades to provide some assistance for generating high-quality software and a smoother development process (citation needed). An overview of the literature shows that most of research has been focusing on the task of predicting both software quality or error appearance (citation needed). As a result, the software life-cycle is often shortened and the maintaining costs reduced. Moreover, by predicting the occurrence of risks, project managers can mitigate delays and reduce (again) the chances of project failures (citation needed). The implications and limitations of the use of this computational techniques are discussed as follows.

The survey of the scientific papers on predicting programming effort has shown that machine learning models are continuously gaining popularity in the academic community. The complexity of applied algorithms is rising as more researchers focus on Deep Learning and continue refining less sophisticated ML models with optimisation algorithms \cite{azzeh2011software}. The obtained results challenge the claims of \cite{jorgensen2014we} that expert estimation is the most reliable method of effort estimation. Instead the study confirms the potential of ML models to provide reliable solutions to SDEE problem, which was first suggested by \cite{srinivasan1995machine} as early as 1995. Empirical evidence of ML models' performance allows the developers to have a greater freedom in selecting various models and tailoring them to a specific project. Subsequently, recent progress in the field encourages more and more publications on the topic. However, when it comes to the direct applications in the industry, these models are not used as frequently as their reported performance would suggest. For instance, among Agile practitioners 63\% use Planning Poker as the primary estimation tool and 38\% prefer expert estimation \cite{usman2015effort}, despite the results of \cite{moharreri2016cost}. The reasons behind this phenomenon are a few limitations of the reviewed scientific papers that hinder the reliability of the results. Due to the lack of large software datasets to use as training data, the studies cannot confirm that their particular results will generalise to every real industrial project. Future studies should make an attempt to gather information about recent software projects, as the majority of currently considered datasets are outdated. 

In third section we have wanted to consider the most popular types of risks related to software development, which we have chosen from \cite{hu2007software}, and decide which of them are more important for development process. This information should be taken into account when considering how to manage software project with minimal losses in the development process. We cannot decide which of these risks are most significant, so, as it was said in \cite{christiansen2015prediction} developers and managers should take into account them all to design really good software project. Because of big difference between considered risks we should use different methods of Machine Learning. Further research is needed to observe a real IT project to find out which of the risks (schedule, budget, technical and management) may affect the development of the project the most negatively. We are also going to find out which risks can be predicted to the maximum extent using Machine Learning.

A substantial amount of research has been conducted with respect to predicting faults and defects using machine learning. The results of the survey conducted show that in predicting faults, machine learning algorithms such as Naive Bayes, K-star, Random Forest and Support Vector Machine have proven to be very beneficial \cite{karim2017software} and more favoured. Moreover, some researchers such as \cite{rashid2014machine} and \cite{rashid2012survey} suggest that Case-based reasoning approach using similarity functions such as Euclidean distance and Manhattan distance to determine the most similar cases, yields encouraging results. While previous research failed to take into consideration the problem of data-set imbalance \cite{shatnawi2012improving}, the outcome of the survey demonstrates that the imbalance has been accommodated. However, it is beyond the scope of this study to specify the  metrics which are relevant in predicting faults. Further research has to make plans for generating new datasets as the available ones, mostly NASA and PROMISE have been used severally.

The overview of the literature shows that some ML techniques, namely case-based reasoning and neural networks, are particularly popular in this field, as shown in Table \ref{reference table}. Case-based reasoning is favoured due to its ability to produce high accuracy given limited data, while neural networks are popular due to their ability to learn complex functions and handle outliers \cite{wen2012systematic}. The reported results build on existing evidence of the usefulness of ML embedded into the software development process. The reliability of such data, however, is affected by the limited available data and the lack for a united and shared dataset.These aspects indicate the need for the development of larger datasets that are representative of current tendencies in software engineering in order to provide researchers with quality training data and allow them to draw reliable conclusions. Future studies should take into account recent developments in the field of ML, such as reinforcement learning, convolutional and recurrent neural networks, providing their applications to software development, which have been scarce to the best of the authors' knowledge.

%% file: 6_Conclusion.tex
\section{Conclusion and future research}
\label{sec:conclusion}

The survey of Machine Learning applications to software development process showcases a considerable progress in the field over last decades. Across three outlined subfields (effort estimation, risks and defects prediction) ML models have been deployed and achieved satisfactory results that are in the majority of cases commensurate to traditional approaches or even surpass them. Literature analysis have also established that increasing research interest in this area provides practitioners a variety of models to apply to their particular project. Given this abundance of models, comparative studies rarely reach consensus about whether traditional regression, classification or deep learning approach is generally preferable in software development.

In the subfield of predicting risks to the software project regression models are considered dominant over other ML models as well as state-of-the-art non-ML methods. Specifically, the performance of Support Vector Machine is frequently noted in regards to predicting schedule and budget risks. On the other hand, defect prediction favours classification algorithms with Random Forest being one of the most reliable models. Research in programming effort estimation initially favoured regression models, however recent breakthroughs confirmed superior accuracy by Cascade Correlation Neural Networks.

Notable gaps in the current state of the research on the topic include investigating larger scope of applications for Artificial Neural Networks and reinforcement learning. Despite that ANNs have shown very promising results in software effort estimation, the research about their applications in two other subfields have been rather scarce. Similar patter is observed regarding reinforcement learning, which was not yet applied to any of the software development tasks mentioned in this paper. 

For future work, it is recommended that researchers attempt to use larger datasets and those that are more representative of the current state of software engineering in order for the models' assessment to be complete and reliable. Moreover, it is advised that closer interaction between academic and industrial communities needs to be established to facilitate deployment of ML models on real-world software projects.

%% file: bibliography.tex
\bibliographystyle{splncs04}
\bibliography{bibliography}